\documentclass[preprint,12pt]{elsarticle}

\usepackage{amssymb}
\usepackage{amsmath}
\usepackage{url}

\journal{Information and Software Technology}

\begin{document}

\begin{frontmatter}

%% Title, authors and addresses

%% use the tnoteref command within \title for footnotes;
%% use the tnotetext command for theassociated footnote;
%% use the fnref command within \author or \affiliation for footnotes;
%% use the fntext command for theassociated footnote;
%% use the corref command within \author for corresponding author footnotes;
%% use the cortext command for theassociated footnote;
%% use the ead command for the email address,
%% and the form \ead[url] for the home page:
%% \title{Title\tnoteref{label1}}
%% \tnotetext[label1]{}
%% \author{Name\corref{cor1}\fnref{label2}}
%% \ead{email address}
%% \ead[url]{home page}
%% \fntext[label2]{}
%% \cortext[cor1]{}
%% \affiliation{organization={},
%%             addressline={},
%%             city={},
%%             postcode={},
%%             state={},
%%             country={}}
%% \fntext[label3]{}

\title{Quantum Executor: A Unified Interface for\\Quantum Computing} %% Article title

%% use optional labels to link authors explicitly to addresses:
%% \author[label1,label2]{}
%% \affiliation[label1]{organization={},
%%             addressline={},
%%             city={},
%%             postcode={},
%%             state={},
%%             country={}}
%%
%% \affiliation[label2]{organization={},
%%             addressline={},
%%             city={},
%%             postcode={},
%%             state={},
%%             country={}}

\author[unipi]{Giuseppe Bisicchia\corref{cor1}}
\cortext[cor1]{Corresponding author.}
\ead{giuseppe.bisicchia@phd.unipi.it}

\author[unipi]{Alessandro Bocci}
\ead{alessandro.bocci@unipi.it}
\author[unipi]{Antonio Brogi}
\ead{antonio.brogi@unipi.it}

%% Author affiliation
\affiliation[unipi]{
organization={Department of Computer Science, University of Pisa},
city={Pisa},
country={Italy}
}

% You are required to provide a concise and factual abstract which does not exceed 300 words.
% Highlights should consist of 3 to 5 bullet points, each a maximum of 85 characters, including spaces.
% Short Communications are limited to 2500 words (approx. 4 pages, a figure counts as 200 words) and should have no more than 10 references.
% Please note that References and appendixes are part of the submission and count against the total number of words.

%% Abstract
\begin{abstract}
%% Text of abstract
As quantum computing evolves from theoretical promise to practical deployment, the demand for robust, portable, and scalable tools for quantum software experimentation is growing. This paper introduces Quantum Executor, a backend-agnostic execution engine designed to orchestrate quantum experiments across heterogeneous platforms. Quantum Executor provides a declarative and modular interface that decouples experiment design from backend execution, enabling seamless interoperability and code reuse across diverse quantum and classical resources. Key features include support for asynchronous and distributed execution, customizable execution strategies and a unified API for managing quantum experiments. We illustrate its applicability through two life-like usage scenarios such as automated benchmarking and hybrid validation, discussing its capacity to streamline quantum development. We conclude by discussing current limitations and outlining a roadmap for future enhancements.
\end{abstract}

% %%Graphical abstract
% \begin{graphicalabstract}
% %\includegraphics{grabs}
% \end{graphicalabstract}

%%Research highlights
% \begin{highlights}
% \item We propose Quantum Executor, a unified quantum orchestration engine.
% \item It enables portable and reproducible quantum experiments across platforms.
% \item It supports asynchronous, distributed, and policy-based experiment execution.

% \end{highlights}

%% Keywords
\begin{keyword}
Quantum Computing \sep Quantum Software Engineering \sep Quantum Software \sep Hybrid Quantum-Cloud Computing 

\end{keyword}

\end{frontmatter}

%% Add \usepackage{lineno} before \begin{document} and uncomment 
%% following line to enable line numbers
%% \linenumbers

%% main text
%%

\section{Introduction}
\label{sec:intro}

Quantum computing is undergoing a remarkable transformation, evolving from a largely theoretical discipline into a field characterized by tangible, programmable devices and accessible cloud-based resources. This rapid progress has created exciting opportunities for researchers and practitioners, who are now able to design, execute, and iterate on quantum experiments with an unprecedented degree of flexibility. As the community embraces these new capabilities, there is a growing need for tools that support robust, reproducible, and scalable quantum software experiments, an essential foundation for both research and real-world deployment~\cite{bisicchia2024quantum}.

A rich ecosystem of high-level software development kits (SDKs) such as Qiskit, Cirq, and Pennylane has emerged to support this transition. These frameworks have been instrumental in lowering the barrier to entry for quantum programming, offering expressive abstractions, comprehensive libraries, and increasingly sophisticated integrations. Their success demonstrates the importance of user-friendly interfaces and modular design principles. At the same time, the proliferation of diverse hardware platforms and provider-specific execution environments has introduced new layers of complexity. As quantum computing platforms continue to diversify and quantum applications become more complex, researchers and developers are often confronted with the challenge of ensuring that their quantum experiments remain portable, interoperable, and maintainable in the face of evolving APIs, runtime conventions, and device architectures.

A few key challenges have become especially prominent:
\begin{itemize}
\item \textbf{Portability:} Moving a quantum experiment between providers typically entails rewriting (parts of) the code to conform to different APIs and object models.
\item \textbf{Interoperability:} Seamlessly integrating resources and coordinating experiments across multiple quantum platforms can be a labor-intensive and technically intricate process.
\item \textbf{Parallelism:} Native support for distributed, asynchronous execution across multiple devices or backends remains limited in most SDKs, hindering support for large-scale or high-throughput experimentation.
% \item \textbf{Reproducibility:} Maintaining transparent and reproducible experiments across changing platforms can be complicated by the tight coupling between applications and provider-specific details.
\end{itemize}

In response to these evolving needs, we present \textit{Quantum Executor}\footnote{\url{https://github.com/GBisi/quantum-executor/}}, a backend-agnostic execution engine purpose-built for orchestrating quantum applications in a multi-provider, multi-device landscape. Rather than replacing existing SDKs, Quantum Executor is designed to complement and extend them. By introducing a higher-level abstraction layer, our approach empowers users to define quantum experiments in a declarative, provider-independent manner, and to leverage asynchronous and distributed execution models with ease.

Key features of Quantum Executor include:
\begin{itemize}
    \item \textit{Declarative experiment definition:} Users can describe quantum experiments independently of the underlying hardware or provider, enhancing portability and code reuse. 
    \item \textit{Transparent orchestration:} Application logic is clearly separated from runtime concerns, enabling robust management of distributed experiments.
    \item \textit{Asynchronous and distributed execution:} Native support for large-scale, parallel experimentation enables the simultaneous submission of multiple quantum experiments across different QPUs. This allows for hybrid quantum-classical workflows, where classical computations can proceed asynchronously while awaiting quantum results, enabling efficient feedback loops and scalable quantum research.
\end{itemize}

Through a unified API and a modular architecture, \textit{Quantum Executor} allows developers to repurpose existing codebases and extend their applications across a range of backends with minimal friction, all while preserving the benefits of established SDKs.

\paragraph{Contributions}
This paper makes the following main contributions:
\begin{itemize}
    \item We introduce \textit{Quantum Executor}, a backend-agnostic orchestration engine that abstracts and unifies quantum experiment execution across multiple hardware and simulator providers.
    \item We propose a modular architecture and declarative API that cleanly separates experiment definition, backend orchestration, and result management, enabling portability, scalability, and reproducibility in quantum software engineering.
    \item We propose a flexible, policy-based orchestration strategy, allowing users to customize experiment distribution and result aggregation through user-defined split and merge policies.
    % \item We discuss realistic, end-to-end usage scenarios that illustrate how \textit{Quantum Executor} can be used in both research and industrial quantum workflows.
    % \item We critically assess current limitations and outline a roadmap for future development to further support robust and scalable quantum experimentation.
\end{itemize}

\noindent
%The remainder of the paper is organized as follows: Section~\ref{sec:related} reviews related work. Section~\ref{sec:library} details the architecture and core features of Quantum Executor. Section~\ref{sec:scenarios} explores practical usage scenarios. Section~\ref{sec:limitations} discusses current limitations, and Section~\ref{sec:conclusions} concludes the paper and offers a roadmap for future development.
\section{Related Work}
\label{sec:related}

% Forest/PyQuil, tket, Orquestra, Strangeworks

Over the past decade, a diverse array of SDKs and frameworks has emerged to support the design, simulation, and execution of quantum circuits. Notable examples include Qiskit~\cite{qiskit}, Cirq\footnote{\url{https://quantumai.google/cirq}}, PennyLane~\cite{pennylane}, and Amazon Braket\footnote{\url{https://amazon-braket-sdk-python.readthedocs.io}}. These platforms have significantly accelerated quantum algorithm development by offering comprehensive tooling for circuit construction, access to quantum backends, and rich pre- and post-processing utilities. However, they are typically optimized for specific provider ecosystems and impose distinct abstractions, which can impede portability and interoperability.

Qiskit, developed by IBM, provides a mature and feature-rich environment for programming IBM Quantum devices, with limited and sometimes fragmented support for non-IBM backends via community extensions. While backend selection abstractions are available, achieving seamless execution across heterogeneous hardware still requires nontrivial code changes and backend-specific adaptations. Cirq and PennyLane, tailored to the Google Quantum and Xanadu ecosystems respectively, also embody provider-centric design philosophies. Although they provide elegant constructs for circuit creation and execution, cross-platform integration often demands bespoke translation logic and limits experiment reuse.

Amazon’s Braket SDK introduces a partial unification by enabling access to multiple hardware providers --- such as IonQ, Rigetti, and OQC --- through a single API. Nevertheless, users are generally constrained to Braket-specific circuit representations, and the integration of circuits from other SDKs remains cumbersome. Support for hybrid and asynchronous experiments is also limited, requiring users to implement orchestration manually.

Among the interoperability-focused initiatives, qBraid\footnote{\url{https://docs.qbraid.com/sdk}} stands out as a notable effort to provide a unified interface for quantum programming across multiple backends and SDKs. qBraid enables users to transpile and execute circuits on a range of hardware and simulators with minimal code changes, and exposes tools for backend management and job submission. However, while qBraid greatly simplifies provider selection and basic code portability, it offers limited support for advanced experiment orchestration, asynchronous or distributed execution, and fine-grained management of results. % In the design of Quantum Executor, we explicitly leverage qBraid as a key building block for backend integration and circuit translation. By incorporating qBraid’s interoperability primitives, Quantum Executor inherits robust circuit portability while extending these capabilities to support declarative experiment definitions, asynchronous dispatch, and fine-grained orchestration across heterogeneous resources.

% Beyond qBraid and other mainstream SDKs, additional projects such as Mitiq~\cite{mitiq}, and OpenQASM~\cite{openqasm} aim to standardize circuit formats, enable noise mitigation across platforms, and support metaprogramming for quantum experiments. These tools provide valuable functionality, but they typically operate within narrow scopes or depend on lower-level integrations, and do not fully address the orchestration and reproducibility challenges posed by multi-provider quantum execution.

\textit{Quantum Executor} aims to advance the state of the art by introducing a backend-agnostic execution layer that complements existing SDKs rather than replacing them. It enables users to express quantum experiments declaratively, decoupling application logic from execution detail, and to dispatch tasks asynchronously across heterogeneous quantum and classical resources. Unlike conventional SDKs, Quantum Executor provides first-class support for distributed execution, real-time management of partial results, and user-defined policies for task scheduling and aggregation. Its modular architecture allows seamless extension to new backends, making it a versatile platform for reproducible, scalable, and provider-independent quantum experimentation.

% In summary, while existing SDKs and tooling ecosystems have laid a strong foundation for quantum software development, they often fall short in addressing the growing need for flexible, portable, and asynchronous experiment orchestration across diverse quantum hardware landscapes. Quantum Executor fills this critical gap by offering a unified interface that abstracts away provider-specific details, while preserving compatibility with established programming paradigms.

\section{Architecture and Implementation}
\label{sec:library}

\textit{Quantum Executor}'s architecture is grounded in a strict separation of concerns: experiment specification, backend orchestration, and result management are realized as distinct but interoperable components. At its core, \textit{Quantum Executor} adopts a provider-agnostic approach, serving as a middleware that decouples high-level experiment definitions from provider-specific details. This modular philosophy not only facilitates the integration of new platforms, but also ensures long-term maintainability as the quantum computing ecosystem continues to evolve.

\begin{figure}
    \centering
    \includegraphics[width=\textwidth]{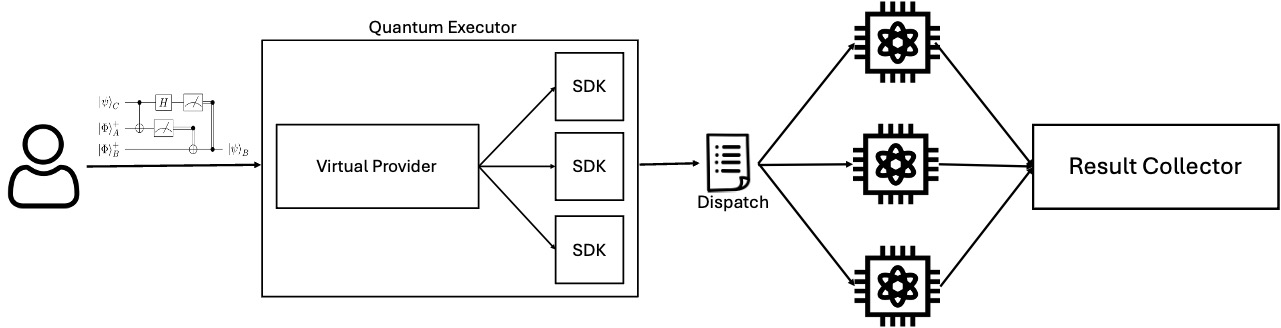}
    \caption{Overview of \textit{Quantum Executor}.}
    \label{fig:overview}
\end{figure}

The system is organized around several key components (see Fig.~\ref{fig:overview}). The \texttt{QuantumExecutor} class acts as the principal orchestrator and user-facing entry point. It handles provider configuration, experiment dispatching, and the selection of execution strategies, exposing a unified API for explicit job control, and declarative or policy-driven execution. Parallelism is natively supported, enabling both synchronous and asynchronous runs, with users able to select multiprocessing for parallel experimentation as needed.

Execution planning is orchestrated through the \texttt{Dispatch} abstraction, which encapsulates the assignment of user-defined circuits to specific quantum providers and backends. Each dispatch defines how circuits (potentially expressed in different intermediate representations) are mapped to appropriate hardware or simulator targets, along with configurable parameters such as shot count and backend-specific options. This abstraction decouples experiment description from low-level execution details, supporting both static and dynamic dispatch strategies. 

Leveraging split policies and a backend-agnostic dispatch mechanism, the Quantum Executor allows users to structure experiments that span multiple circuits and backends without duplicating logic, significantly simplifying the design of complex and heterogeneous quantum workflows.

Result management is centralized in the \texttt{ResultCollector}, which monitors the lifecycle of submitted jobs and aggregates their outputs. This component supports both blocking and non-blocking retrieval, making it suitable for both interactive sessions and high-throughput pipelines. Aggregation of results is policy-driven, with support for both default and user-defined merge strategies. Results can be accessed in structured formats, including tabular representations for downstream analysis.

A notable architectural feature is the policy system, adapted from our previous works, e.g.,~\cite{bisicchia2023distributing, bisicchia2023dispatching}, which governs both experiment distribution (via split policies) and result aggregation (via merge policies). Policies are implemented as Python functions that may be registered at runtime, allowing users to encode domain-specific logic or advanced orchestration strategies without modifying the library core. For example, a split policy might partition an experiment's shots evenly across all available backends, while a merge policy could implement sophisticated post-processing or consensus mechanisms.

Central to the library’s extensibility is the \texttt{VirtualProvider} abstraction. This component serves as a unifying interface to heterogeneous provider SDKs (including Qiskit, Cirq, PennyLane, and Braket) handling backend discovery, credential management, and provider-specific quirks behind a common API. \texttt{VirtualProvider} leverages qBraid primitives, ensuring seamless interoperability even as provider APIs diverge.

The execution flow in Quantum Executor begins with the user defining an experiment, either programmatically or in a declarative fashion. The \texttt{QuantumExecutor} constructs a \texttt{Dispatch} object, directly or by invoking a split policy, and distributes jobs across the selected providers and backends. The \texttt{ResultCollector} monitors execution, enabling asynchronous result retrieval when desired, and applies the specified merge policy to produce coherent experiment-level outputs. All results are ultimately presented through a unified interface that abstracts over provider differences, facilitating immediate inspection or further analysis.

% Multiprocessing and thread-safe constructs underpin parallel execution, while the modular design ensures that new providers, policies, and aggregation strategies can be integrated with minimal friction. 

% This architecture enables reproducible, scalable, and provider-independent quantum experimentation, positioning Quantum Executor as a robust foundation for research and real-world deployment in a rapidly changing landscape.
\section{Usage Scenarios}
\label{sec:scenarios}

To illustrate the features of \textit{Quantum Executor}, we describe two realistic usage scenarios inspired by industrial and research needs. These examples highlight the platform's ability to support robust, portable, and declarative quantum experimentation across heterogeneous resources.

\subsection{Scenario 1: Automated Batch Benchmarking on All Available Backends}

%\paragraph{Context}
A research team aims to periodically benchmark a reference workload on every accessible quantum device, to monitor hardware performance, detect regressions, and ensure system health over time. The benchmarking employs the \texttt{multiplier} split policy to ensure each experiment is executed the same exact number of times (shots) on all backends.

% \paragraph{Objectives}
% \begin{itemize}
%     \item Automate dispatch of a set of circuits to every available QPU with identical parameters.
%     \item Use the \texttt{multiplier} split policy to ensure each experiment is executed the same number of times on all platforms.
%     \item Collect results in parallel and archive them for further analysis and dashboarding.
% \end{itemize}

% \paragraph{Sketched Implementation}

\begin{center}
\begin{minipage}{\linewidth}
\begin{verbatim}
from quantum_executor import QuantumExecutor

circuits = [qc1, qc2, qc3, ...] 

qe = QuantumExecutor(providers_info={
    "ionq": {"api_key": "⟨IONQ_API_KEY⟩"},
    "qbraid": {"api_key": "⟨QBRAID_API_KEY⟩"},
})
all_backends = qe.virtual_provider.get_backends(online=True)

collector = qe.run_experiment(
    circuits=circuits,
    shots=1024,
    backends={prov: list(backs.keys()) for prov, backs in all_backends.items()},
    split_policy="multiplier",
    multiprocess=True,
    wait=True
)
d
results = collector.get_results()
# Example: {'ionq': {'qpu.aria-1': {'0000': 123, '0110':13, ...
\end{verbatim}
\end{minipage}%
\end{center}

%\paragraph{Expected Outcome}
\begin{itemize}
    \item \textbf{Scalability:}  The same set of circuits is automatically dispatched and executed across all available backends, showcasing Quantum Executor ability to orchestrate multi-backend experiments with minimal configuration effort.
    \item \textbf{Maintainability:} No code changes are needed as providers and hardware evolve.
    \item \textbf{Transparency:} Results are consistently structured, supporting automated monitoring and analytics pipelines.
\end{itemize}

\subsection{Scenario 2: Enterprise Benchmarking—Simulated vs. Real Quantum Execution}

%\paragraph{Context}
A corporate R\&D team integrates a quantum kernel into a business-critical workflow, such as supply chain optimization or portfolio rebalancing. To ensure ongoing correctness and monitor hardware performance, the team routinely benchmarks their circuit both on a noise-free simulator and on selected quantum hardware. Comparing these results allows detection of fidelity regressions and enables informed provider selection. This evaluation employs a custom merge policy that computes the Total Variation Distance (TVD) between each QPU’s output and the ideal simulator output, quantifying the deviation introduced by hardware noise.

% \paragraph{Objectives}
% \begin{itemize}
%     \item Seamlessly execute the same circuit on a perfect simulator and on real quantum hardware backends.
%     \item Run jobs in parallel, collecting results in a uniform structure.
%     \item Apply a custom merge policy that computes the Total Variation Distance (TVD) between each QPU's output distribution and the simulator distribution.
%     \item Automate result retrieval for easy archival and comparison.
% \end{itemize}

%\paragraph{Sketched Implementation}

\begin{center}
\begin{minipage}{\linewidth}
\begin{verbatim}
from quantum_executor import QuantumExecutor

# Merge policy: returns TVD of each QPU from simulator
def tvd_merge(results, _):
    sim = results["local_aer"]["aer_simulator"][0]
    tvd_dict = {}
    for prov, backs in results.items():
        for bname, runs in backs.items():
            if prov == "local_aer": continue
            tvd = total_variation_distance(runs[0], sim)
            tvd_dict[f"{prov}/{bname}"] = tvd
    return tvd_dict, {}

qc = ... 

qe = QuantumExecutor(
    providers_info={...},
    providers=["local_aer", "ionq", "qbraid"]
)
qe.add_policy(name="tvd", merge_policy=tvd_merge)

backends = {
    "local_aer": ["aer_simulator"],
    "ionq": ["qpu.forte-1"],
    "qbraid": ["ibm_toronto"]
}

collector = qe.run_experiment(
    circuits=qc,
    shots=2048,
    backends=backends,
    split_policy="multiplier",
    merge_policy="tvd",
    multiprocess=True,
    wait=True
)

tvd_results = collector.get_merged_results()
# Example: {'ionq/qpu.forte-1': 0.18, 'qbraid/ibm_toronto': 0.32}
\end{verbatim}
\end{minipage}%
\end{center}

%\paragraph{Expected Outcome}
\begin{itemize}
    \item \textbf{Portability:} The same circuit and invocation logic is executed across simulator and hardware with no code changes. If the company decides to change the target QPU, the execution logic remains exactly the same.
    \item \textbf{Parallelism:} Jobs are dispatched in parallel, reducing turnaround time.
    \item \textbf{Quantitative Comparison:} The custom merge policy delivers a TVD score for each QPU, quantifying divergence from the perfect (simulated) distribution and enabling clear provider benchmarking.
\end{itemize}

% These scenarios illustrate how \textit{Quantum Executor} may enable robust, reproducible, and scalable quantum experimentation in real‐world, multi‐provider environments, streamlining workflows for both enterprise and research teams.
\section{Known Limitations}
\label{sec:limitations}

While \textit{Quantum Executor} offers a unified and extensible interface for orchestrating quantum experiments across diverse providers, it is important to acknowledge its current limitations to contextualize its appropriate use and guide future improvements.

\begin{itemize}
    % \item \textbf{Backend Diversity and API Drift.} The rapid evolution of quantum hardware platforms and provider SDKs means that API changes or new backend features may not be immediately supported. Despite the provider-agnostic design, users may still encounter incompatibilities or require manual intervention when novel devices or significant provider updates are introduced.

    % \item \textbf{Limited Native Error Mitigation.} While \textit{Quantum Executor} supports the integration of custom post-processing and aggregation policies (e.g., noise-aware merging), it does not provide native, built-in mechanisms for error mitigation, calibration, or advanced device-specific optimizations. Leveraging such features requires additional user logic or external toolkits.

    % \item \textbf{Circuit and Data Format Assumptions.} The library currently assumes a certain degree of compatibility in circuit and result data formats between supported SDKs (such as Qiskit, Cirq, and qBraid). Highly specialized or provider-specific quantum operations may require custom translation layers, which are not handled automatically.

    \item \textbf{Resource Management and Scalability.} Although parallel and distributed execution is supported, large-scale experimentation remains constrained by the resource limits of underlying hardware and provider job queues. There is no automatic load balancing or job scheduling beyond the selected split policies; scalability may be affected by backend-specific throttling or access policies.

    %\item \textbf{Security and Credential Management.} Quantum Executor relies on user-supplied credentials for interacting with third-party providers. Secure management, rotation, and revocation of these credentials are outside the current scope, and must be handled by the user according to organizational best practices.

    \item \textbf{Hybrid Quantum-Classical Workflows.} While the system is suitable for quantum job orchestration, advanced hybrid algorithms (e.g., variational or feedback-based workflows) require additional coordination logic outside the current abstraction, and may benefit from tighter integration with classical orchestration platforms.
\end{itemize}

By clearly stating these limitations, our aim is to provide transparency regarding the intended scope of \textit{Quantum Executor}, suggest several directions for future work and motivate further community-driven enhancements that will progressively overcome these constraints in future versions, aiming to foster an open and extensible quantum orchestration ecosystem.
\section{Conclusions}
\label{sec:conclusions}

This work introduced \textit{Quantum Executor}, a unified, backend-agnostic execution engine for orchestrating quantum experiments across heterogeneous quantum devices and simulators. By decoupling experiment specification from backend orchestration, \textit{Quantum Executor} enables researchers and practitioners to design, execute, and analyze quantum workflows with enhanced portability, reproducibility, and scalability. Through a unified API, flexible policy-based orchestration, and support for both synchronous and asynchronous execution, our approach streamlines the integration of quantum resources into both research and industrial pipelines.

% \subsection*{Roadmap and Future Work}

% Building on the current implementation, several directions are envisioned to further enhance \textit{Quantum Executor}:

% \begin{itemize}
%     % \item \textbf{Native Error Mitigation and Calibration.} Integration with state-of-the-art error mitigation, noise-aware scheduling, and calibration routines to improve experiment reliability on near-term devices.
%     \item \textbf{Hybrid Workflow Support.} Native support for hybrid quantum-classical algorithms, tightly integrated with classical orchestration engines.
%     \item \textbf{Dynamic Resource Management.} Advanced features to optimize for performance and cost.
%     %\item \textbf{Security Enhancements.} Improved credential management, secure execution modes, and integration with enterprise authentication systems.
%     \item \textbf{Community and Ecosystem Expansion.} Encouraging contributions to foster an open and extensible quantum orchestration ecosystem.
% \end{itemize}

% These improvements aim to further streamline quantum software experimentation and accelerate the transition from prototype to production in heterogeneous, multi-provider environments.

%% If you have bib database file and want bibtex to generate the
%% bibitems, please use
%%
%%  \bibliographystyle{elsarticle-num-names} 
%%  \bibliography{<your bibdatabase>}

%% else use the following coding to input the bibitems directly in the
%% TeX file.

%% Refer following link for more details about bibliography and citations.
%% https://en.wikibooks.org/wiki/LaTeX/Bibliography_Management

\bibliographystyle{elsarticle-num-names} 
\bibliography{biblio}

\end{document}